\documentclass[pra,twocolumn,aps,nofootinbib,superscriptaddress,longbibliography]{revtex4-2}

\usepackage{times}
\usepackage{amsmath}
\usepackage{amssymb}
\usepackage{graphicx}
\usepackage{braket}
\usepackage{hyperref}

\hypersetup{
  colorlinks=true,
  citecolor=blue,
  linkcolor=blue,
  urlcolor=blue
}

\DeclareMathOperator{\sgn}{sgn}
\newcommand{\Tr}[1]{\operatorname{Tr}\{#1\}}

\begin{document}
\title{Homogeneous Magnetic Flux in Rydberg Lattices}
\author{J.~Eix$^+$}
\email{jeix@iastate.edu}
\affiliation{Institute for Theoretical Physics III and Center for Integrated Quantum Science and Technology, Universität Stuttgart, Pfaffenwaldring 57, 70569 Stuttgart, Germany}
\affiliation{Department of Physics and Astronomy, Iowa State University, Ames, Iowa 50011, USA}
\author{R.~Bai$^+$}
\email{rukmani.bai@itp3.uni-stuttgart.de}
\affiliation{Institute for Theoretical Physics III and Center for Integrated Quantum Science and Technology, Universität Stuttgart, Pfaffenwaldring 57, 70569 Stuttgart, Germany}
\author{T.~Lahaye}
\affiliation{Universit\'e Paris-Saclay, Institut d’Optique Graduate School, CNRS, Laboratoire Charles Fabry, 91127 Palaiseau Cedex, France}
\author{A.~Browaeys}
\affiliation{Universit\'e Paris-Saclay, Institut d’Optique Graduate School, CNRS, Laboratoire Charles Fabry, 91127 Palaiseau Cedex, France}
\author{H.~P.~B\"uchler}
\affiliation{Institute for Theoretical Physics III and Center for Integrated Quantum Science and Technology, Universität Stuttgart, Pfaffenwaldring 57, 70569 Stuttgart, Germany}
\author{S.~Weber}
\email{sebastian.weber@itp3.uni-stuttgart.de}
\affiliation{Institute for Theoretical Physics III and Center for Integrated Quantum Science and Technology, Universität Stuttgart, Pfaffenwaldring 57, 70569 Stuttgart, Germany}

\def\thefootnote{+}\footnotetext{These authors contributed equally to this work.}

\begin{abstract}
We present a method for generating homogeneous and tunable magnetic flux for bosonic particles in a lattice using Rydberg atoms. Our setup relies on Rydberg excitations hopping through the lattice by dipolar exchange interactions. The magnetic flux arises from complex hopping via ancilla atoms. Remarkably, the total flux within a magnetic unit cell directly depends on the ratio of the number of lattice sites to ancilla atoms, making it topologically protected to small changes in the positions of the atoms. This allows us to optimize the positions of the ancilla atoms to make the flux through the magnetic unit cell homogeneous. With this homogeneous flux, we get a topological flat band in the single-particle regime. In the many-body regime, we obtain indications of a bosonic fractional Chern insulator state at $\nu = 1/2$ filling.
\end{abstract}

\maketitle

\section{Introduction}
Charged particles in the presence of magnetic fields exhibit many fascinating phenomena \cite{hofstadter_76, laughlin_83, Nijs_82, Halperin_82, simon_83, kohmoto_85, Hatsugai_93} including integer \cite{klitzing_80} and fractional \cite{stormer_82} quantum Hall states, as well as topological fractional Chern insulating states \cite{regnault_11}. The observation of these many-body states requires strong interactions between the charged particles and high magnetic fields. In this context, quantum simulators comprised of ultracold atoms trapped in optical lattices have emerged as a pathway to achieve these conditions, owing to their many tunable experimental parameters \cite{bloch_12, gross_17, florian_20, jaksch_98, greiner_01, greiner_02}.  
However, ultracold atoms are charge neutral and do not experience the Lorentz force when exposed to external magnetic fields.
Despite this obstacle, magnetic fields for neutral atoms can be implemented artificially \cite{jaksch_03, delibard_11, lin_09b, lin_09a, lin_11, aidelsburger_11, miyake_13}. There are several ways to generate effective magnetic fields, including rotating the optical lattice \cite{williams_10}, using laser assisted tunneling \cite{jaksch_03, aidelsburger_11, miyake_13}, implementing spatially inhomogeneous light fields \cite{lin_09b, lin_09a, lin_11}, periodically perturbing the Hamiltonian \cite{jimenez_12, struck_12}, and lattice shaking \cite{price_17,delibard_11}. In this work, we propose an alternative method to generate homogeneous magnetic flux in rectangular lattices of charge neutral particles using only lattice geometry and Rydberg atoms.

Rydberg atoms trapped in optical tweezer arrays provide a convenient platform to study many body physics for charge neutral particles in effective magnetic fields due to the experimental control of their long-range dipole-dipole interactions \cite{deleleuc_18}, the tunable geometry of the arranged atoms, and dipolar exchange interactions \cite{saffman_10, weimer_10}. Effective magnetic fields arise from complex hopping, known as the Peierls phase \cite{peierls_33}, which can be realized solely from dipolar exchange interactions in Rydberg atoms without requiring inhomogeneous light fields or lattice shaking \cite{lienhard_20}. Rydberg atom arrays have been identified as a tool to explore different quantum states of matter such as symmetry-protected topological phases \cite{peter_15, deleleuc_19, weber_18}, fractional Chern insulating states \cite{weber_22}, and quantum spin liquid states \cite{semeghini_21}. However, it remains an open challenge to find setups where the effective magnetic field corresponds to a homogeneous magnetic flux.

Here, we provide a way to generate robust, homogeneous, and tunable magnetic flux in rectangular lattices of Rydberg atoms. Our method makes use of ancilla atoms in a background lattice which induce an effective magnetic flux arising from dipolar exchange interactions. Strikingly, the total flux per magnetic unit cell is determined by the ratio of ancilla atoms to Rydberg atoms. This ratio is a bulk characteristic of the system and robust to perturbation; hence, the total flux through the magnetic unit cell is a topological property, but the flux can be tuned locally by varying the lattice geometry. Using this method, flat topological bands can be generated with arbitrary flux values. After establishing the robustness of the magnetic flux, we find indications for a bosonic fractional Chern insulator state at $\nu = 1/2$ filling in the many body regime, in agreement with previous studies that theoretically predicted such a state in a bosonic Harper Hofstadter model \cite{bai_18, hafezi_07, palmer_06, wang_11, gerster_17}.

This article is organized as follows. In section II, we discuss our setup along with the Rydberg level scheme. In section III, we examine obtainable magnetic flux patterns and show that the total flux through the magnetic unit cell is determined by the ratio of sites between the background and rectangular lattices. Furthermore, we demonstrate how the magnetic flux can be made homogeneous by optimizing the lattice. In section IV, we discuss the topological flat bands in the single-particle picture. Finally, in section V, we investigate the many-body regime and obtain a topological state at $\nu = 1/2$ filling using the robust flux generated from our method.

\section{Setup}
\label{sec2}
\begin{figure}
    \centering
    \includegraphics[width=\columnwidth]{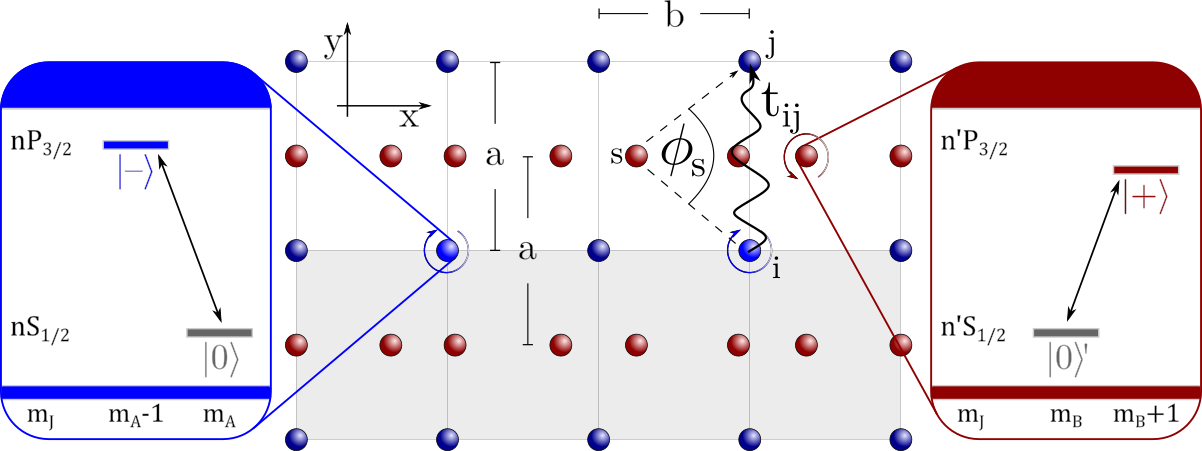}
    \caption{Setup with rectangular lattice ${\rm A}$ (blue sites) hosting $\ket{-}$ excitations which encode the hopping model. The background lattice ${\rm B}$ (red sites) of ancilla atoms hosts $\ket{+}$ excitations and generates complex hopping in lattice ${\rm A}$. Level structures of lattice ${\rm A}$ and ${\rm B}$ sites are shown. Additionally depicted is an exemplary second-order process, which contributes a complex phase based on the angle $\phi_{isj}$ to the hopping $t^{-}_{ij}$, as described in Eq. (\ref{Eq1}). The magnetic unit cell is shaded in gray.}
    \label{Fig1}
\end{figure}

We begin with the microscopic description of our setup. We focus on Rydberg atoms 
trapped individually by optical tweezers, where each atom can occupy one of two Rydberg states which are energetically isolated from all other Rydberg states. The atoms are arranged to form a rectangular lattice $\rm A$ (blue sites) and background lattice $\rm B$ of ancilla atoms (red sites), as depicted in Fig.\ref{Fig1}.

We start by describing lattice ${\rm A}$ in detail. Lattice ${\rm A}$ has horizontal spacing $b$ and vertical spacing $a$. Rydberg atoms in lattice ${\rm A}$ have quantum number $n$ and an internal level structure consisting of two different Rydberg states, namely a Rydberg S-state $\ket{0}$ and a Rydberg P-state $\ket{-}$, as shown in Fig.\ref{Fig1} \cite{weber_18}. The states can be isolated energetically from other states using, for example, a combination of electric and magnetic fields \cite{lienhard_20}.
The state $\ket{0}$ can be interpreted as the vacuum state, and the state $\ket{-}$ can be interpreted as a bosonic excitation. Therefore, the model is described by hard-core bosons with bosonic operators satisfying $b^{\dagger}_{-}\ket{0} = \ket{-}$ \cite{peter_15}. Within the lattice, dipolar exchange interactions give rise to hopping of $\ket{-}$ excitations, which can in general be described by the Hamiltonian
\begin{align}
\label{haml}
\mathcal{H} =
\sum_{ij}t^{-}_{ij}b^{\dagger}_{-,j}b_{-,i} \;,
\end{align}
where $t^{-}_{ij}$ is the hopping element
between site $i$ and $j$ of lattice $\rm A$.

We use the ancilla lattice ${\rm B}$ to generate effective magnetic fields. We do so by allowing off-resonant hopping processes that occupy the ancilla atoms virtually, making the hopping elements $t^{-}_{ij}$ complex. To achieve these hopping processes, lattice ${\rm B}$ hosts Rydberg atoms with quantum number $n'$ and an internal level structure consisting of a Rydberg S-state $\ket{0}'$ and a Rydberg P-state $\ket{+}$, as shown in Fig. \ref{Fig1}. Having two different quantum numbers  $n$ and $n'$ for the atoms forming the rectangular lattice ${\rm A}$ and the background lattice ${\rm B}$, allows us to choose level schemes where virtual hopping processes among atoms within the same lattice are negligible. Note that we specifically chose the $\ket{-}$ and $\ket{+}$ state to differ in the magnetic quantum number by two. All lattice ${\rm B}$ sites are initialized in the ground state $\ket{0}'$. To ensure that the lattice ${\rm B}$ sites are only occupied virtually, we apply an energy shift $\mu = \left[E(\ket{+})-E(\ket{0}') - E(\ket{-})+E(\ket{0})\right]$. This shift allows us to treat the hopping from a site $i$ to a site $j$ through an ancilla atom in second-order perturbation theory. We have illustrated such a second-order process in Fig.\ref{Fig1}, where a $\ket{-}$ excitation at site $i$ hops to site $j$ by first virtually occupying the ancilla atom $s$ with a $\ket{+}$ excitation. The spin-flip hopping processes couple the $\ket{+}$ to the $\ket{-}$ state, changing the internal angular momentum by two which yields a complex phase due to the conservation of angular momentum \cite{lienhard_20}. 

Taking into account these processes, the hopping element $t^{-}_{ij}$ becomes
\begin{equation}
  t^{-}_{ij} = \vert{t^{-}_{ij}}\vert e^{i\phi_{ij}} 
	= -\frac{t^-}{{\vert{\mathbf{R}_{ij}}\vert}^3} 
	- \sum_s \frac{w^2}{\mu{\vert{\mathbf{R}_{is}}\vert}^3
	{\vert{\mathbf{R}_{sj}}\vert}^3}e^{2i\phi_{isj}}\;,
    \label{Eq1}
\end{equation}
where the first term is the direct hopping from $i$ to $j$ with hopping strength $t^{-}$, and the second term arises from second-order hopping processes. The phase factor of a second-order process is determined by the in-plane azimuthal angle $\phi_{isj}$ shown in Fig. \ref{Fig1}. The sign of $\phi_{isj}$ is determined by $\sgn{\mathbf{R}_{ij}\times\mathbf{R}_{is}}$ where $\mathbf{R}_{ij}$ is the vector from $i$ to $j$. The amplitude of a second-order process is determined by the strength $w$ of the hopping between lattice ${\rm A}$ and ${\rm B}$ and the energy difference $\mu$ \cite{peter_15, weber_18}. Because the value of $t^{-}/w$ is highly implementation dependent and the results of our paper stay qualitatively the same for different values of $t^{-}$, we consider the limit $w\gg t^{-}$ and set $t^{-}=0$ for the rest of this work.

The ancilla lattice ${\rm B}$ is chosen such that it has the same periodicity in the $y$ direction as lattice ${\rm A}$. The periodicity of lattice $\rm B$ in the $x$ direction is commensurable with the periodicity of lattice ${\rm A}$. This allows us to specify a unit cell of the combined lattice with $p$ lattice ${\rm B}$ sites and $q$ lattice ${\rm A}$ sites, being invariant under translations by $qb\hat{x}$ and $a\hat{y}$. For an example of a unit cell with $q=4$, $p=7$, see the gray shaded region in Fig. \ref{Fig1}. This unit cell coincides with the magnetic unit cell of the lattice ${\rm A}$. We align lattice ${\rm B}$ so that, in the $y$ direction, sites of lattice ${\rm A}$ are centered between sites of lattice ${\rm B}$. This guarantees real hopping elements along $x$ since pairs of lattice ${\rm B}$ sites contribute to the second-order hopping with conjugate phases 
\cite{footnote}.The integer-valued number of flux quanta through the magnetic unit cell is fully determined by the complex hoppings along $y$. Remarkably, as we will show in the next section, the number of flux quanta is directly related to the numbers $q$ and $p$ of lattice sites in the magnetic unit cell.
\begin{figure}
    \centering
    \includegraphics[width=\columnwidth]{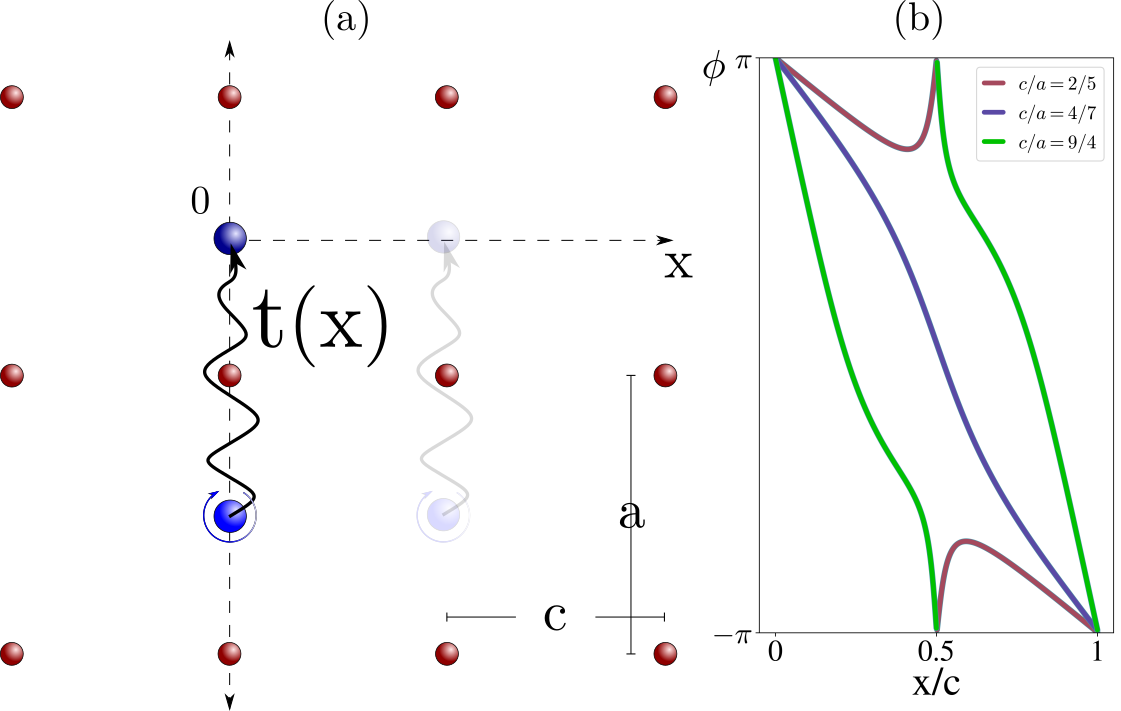}
    \caption{(a) Two lattice ${\rm A}$ sites immersed in lattice ${\rm B}$, which is held to be a rectangular lattice. (b) Upon smoothly translating the two sites by the horizontal lattice constant $c$, the phase angle $\phi$ of the complex hopping $t(x)$ collects a phase $2\pi n$ where $n$ is the winding number. The winding number depends on the ratio $c/a$. It is zero/one/two for the red/purple/green curve, respectively.}
    \label{Fig2}
\end{figure}

\section{Generating Homogeneous Magnetic Fields}
\label{sec3}

In this section, we show that the effective magnetic flux can be tuned to be homogeneous, and arbitrary rational numbers of flux quanta can be realized in the Rydberg lattice ${\rm A}$. As explained in the previous section \ref{sec2}, only the hopping along $y$ direction is complex-valued. To characterize this hopping, we consider two sites of lattice ${\rm A}$ within an infinite lattice ${\rm B}$, as shown in Fig. \ref{Fig2} (a). For our argument and without loss of generality, lattice ${\rm B}$ is taken to be rectangular with horizontal lattice spacing $c$ and vertical spacing $a$. The two lattice ${\rm A}$ sites are a pair of vertical nearest-neighbors separated by $a$. 

While our discussion holds for arbitrary ratios of $c/a$, we consider $c = 4/7a$ for illustration, being consistent with the setup depicted in Fig. \ref{Fig1} that realizes this value for $q=4$, $p=7$, and $a=b$. In the following, we move the pair of lattice ${\rm A}$ sites along $x$ while keeping track of the complex hopping between them. The hopping is thus a function of horizontal distance $x$ and is denoted as $t(x)$. The phase of $t(x)$ is denoted as $\phi(x)$.

The phase $\phi(x)$ is plotted in Fig. \ref{Fig2} (b) as the pair of lattice ${\rm A}$ sites are translated along $x$ from $0$ to $c$. This translation reveals a few general and key properties of $\phi(x)$. First, due to translational symmetry, $t(x)$ is invariant when $x$ changes by an integer multiple of $c$. Accordingly, $\phi(x)$ is invariant with $\phi(x+c)=\phi(x)\mod 2\pi$. Second, we observe that $\phi(mc+x) = -\phi(mc-x)~ \forall \, m \in\frac{1}{2}\mathbb{Z}$ because the corresponding geometrical arrangements are related by a reflection. The result here can be understood from the fact that $t_{ij} = t_{ji}^{*}$ from Eq. (\ref{Eq1}), and that a reflection effectively reverses the hopping direction. Third, the phase $\phi(x)$ is a continuous function in the interval $[0,c]$ as long as $t(x) \neq 0$.
Importantly, these properties imply that $\phi(x)$ can only change by $2\pi n$ when $x$ increases from $0$ to $c$. We call the integer value $n$ the winding number.

It turns out that the winding number is dependent on the ratio $c/a$. For example, for $c/a = 4/7$, we observe that $\phi(x)$ varies from $\pi$ to $-\pi$ as $x$ varies from $0$ to $c$, giving a winding number of $1$. In contrast, for $c/a \lesssim 0.41$, the winding number is found to be $0$. For $c/a \gtrsim 2.00$, the winding number is $2$. In between, $0.41 \lesssim c/a \lesssim 2.00$, the winding number is $1$, see Fig. \ref{Fig2} (b). The values $c/a \approx 0.41$ and $c/a \approx 2.00$ are critical values which separate different regimes. These critical values correspond to special cases in which the hopping $t(x)$ vanishes at the reflection symmetric point $x=c/2$; thus, the phase $\phi(c/2)$ and the winding number are no longer defined. 

\begin{figure}
    \centering
    \includegraphics[width=1.0\columnwidth]{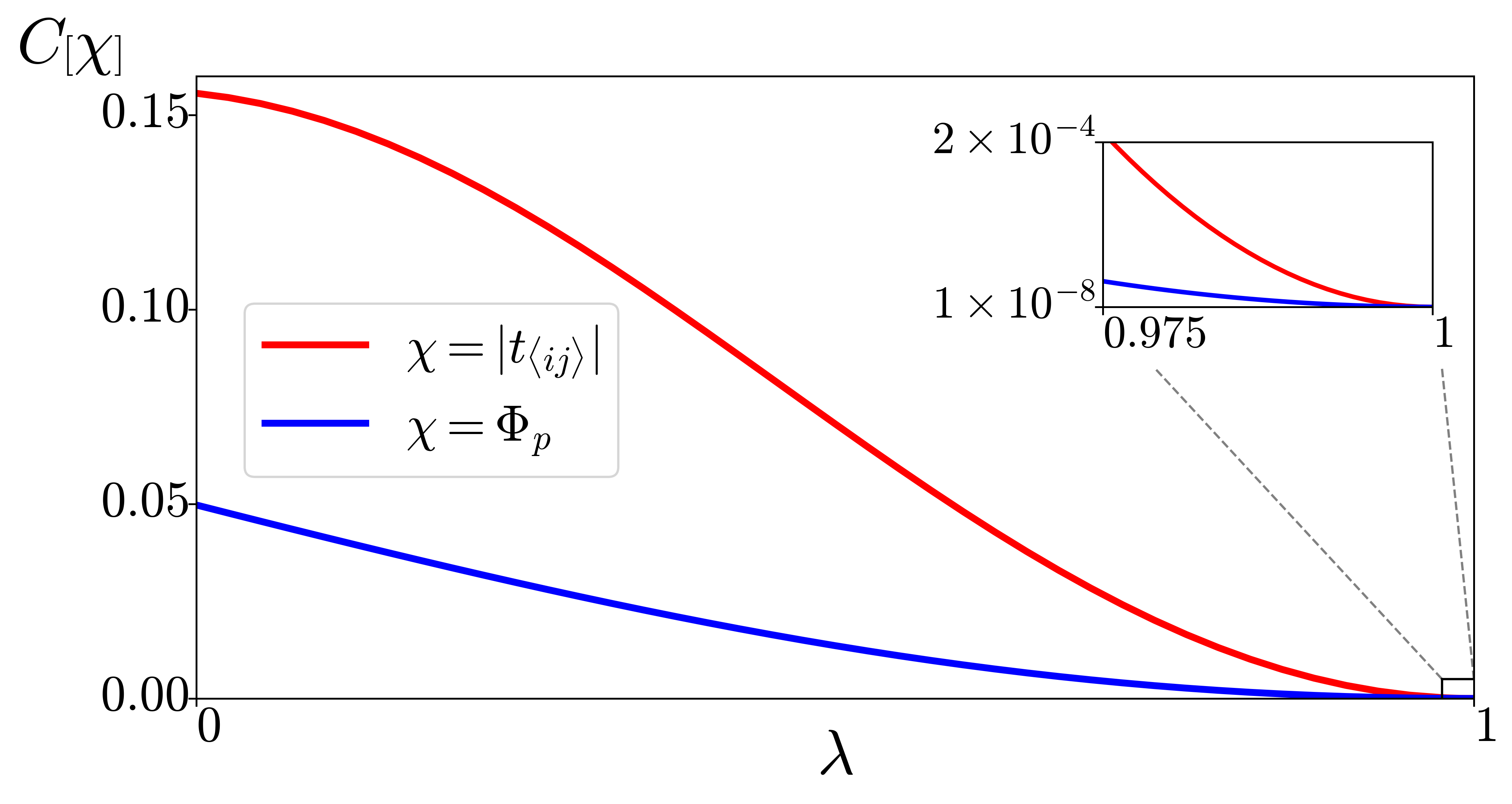}
    \caption{Optimization for homogeneous flux and hopping amplitude. Normalized variance $C[\chi] \equiv \frac{\mathbb{E}[\chi^2] - \mathbb{E}[\chi]^2}{\mathbb{E}[\chi]^2}$ (where $\mathbb{E}[\chi]$ is the average over the whole lattice) of nearest-neighbor hopping amplitude $\vert t_{\langle ij\rangle}\vert$ and plaquette flux $\Phi_p$ vs. the linear interpolation parameter $\lambda$. $\lambda=0$ corresponds to the initialized lattice, while $\lambda=1$ corresponds to the fully optimized lattice. The mean flux is always $\pi/4$ while the mean of the hopping amplitude goes from $33.10 \frac{w^2}{\mu b^6}$ to $36.59 \frac{w^2}{\mu b^6}$. The inset shows the variance for both the hopping and the flux trends to zero.}
    \label{FigOpt}
\end{figure}

This behavior of the phase $\phi(x)$ has implications for the flux through a magnetic unit cell. Let us consider that the lattice ${\rm A}$ has a rational horizontal spacing $b = \frac{p}{q}c$. Then, the magnetic unit cell will contain $p$ lattice ${\rm B}$ sites and $q$ lattice ${\rm A}$ sites. Our analysis of the winding number showed that the phase $\phi(x)$ of the hopping along $y$ increments by $2\pi n$ with  $n \in \{0,1,2\}$ between each lattice ${\rm B}$ site. Accordingly, the total flux through the magnetic unit cell must be $2\pi pn$ because we have $p$ lattice ${\rm B}$ sites. The average flux through each lattice ${\rm A}$ plaquette is thus $2\pi pn/q$.

\begin{figure*}[ht]
    \centering
    \includegraphics[width=2\columnwidth]{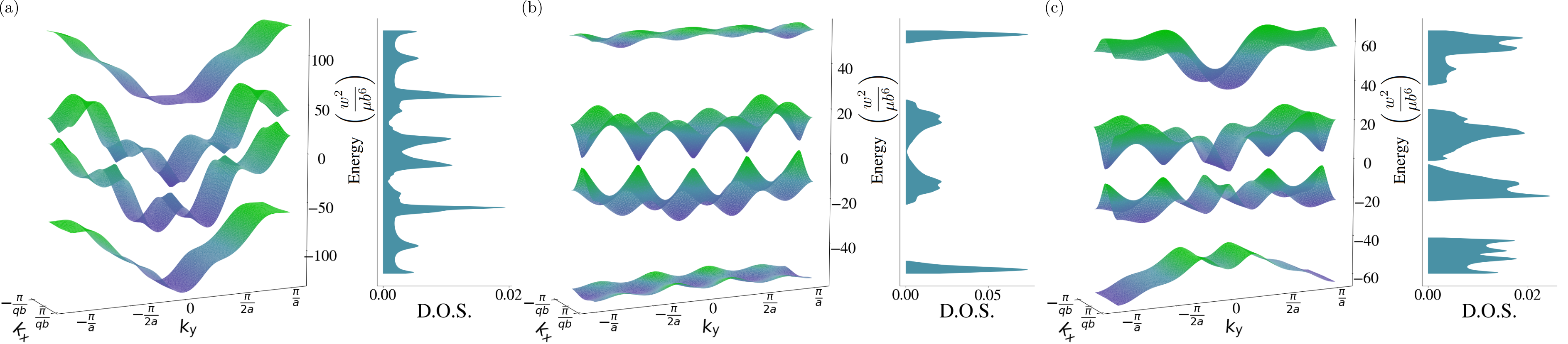}
    \caption{Band structures and densities of states for magnetic unit cells comprised of $q=4$ lattice ${\rm A}$ sites with a background lattice ${\rm B}$ of $p=7$ sites. (a) Nearest neighbor hoppings on the unoptimized lattice. (b) Nearest neighbor hoppings on the optimized lattice. The band structure is clearly flattened, and the density of states shows the opening of a band gap that is on the order of the mean of the nearest-neighbor hopping amplitudes. (c) Long range hoppings on the optimized lattice. Although longer range hoppings roughen the bands, the bands remain gapped.}
    \label{Fig3}
\end{figure*}

Notably, our argument did not rely on specific positions of the lattice sites. Thus, rearranging the positions of the sites in the magnetic unit cell does not change the total flux through it as long as $t(x)$ does not vanish. This allows us to optimize the homogeneity of the flux and hopping strengths by adapting the positions of the lattice sites. 

While tuning the positions of each site is a high-dimensional optimization problem that is challenging to solve exactly, we use a heuristic that provides reasonably homogeneous magnetic fields and hoppings. Our heuristic takes place in several stages. At the beginning, lattice ${\rm A}$ is initialized with $a=b$, and the background lattice ${\rm B}$ is initialized with vertical spacing $a$ and horizontal lattice spacing $c = (q/p)a$. This fixes the average flux to $2\pi p/q$, provided that $0.41 \lesssim q/p \lesssim 2.00$. As a first optimization step, the flux is optimized for homogeneity by rearranging lattice $\rm{B}$ sites within the magnetic unit cell along $x$. During this process, we keep the magnetic unit cell symmetric under reflection. As a second optimization step, we optimize for homogeneous hopping strengths. To do so, we tune the ratio between $a$ and $b$ and, in addition, allow lattice $\rm B$ sites to move out of the plane in $z$ direction. Again, we only allow for rearrangements that keep the reflection symmetry. Optimizing for homogeneous flux or hopping strength are not independent processes; however, iterating the processes converges in that flux variance and hopping strength variance may both be minimized, as shown in Fig. \ref{FigOpt}. As we will see in the next section, having a homogeneous flux and nearly homogeneous hoppings allow for obtaining topological flat bands in our model.

\section{Topological Band Structure}
In this section, we present the topological band structure of lattice ${\rm A}$ for the optimized and not optimized positions of lattice ${\rm B}$ sites.
In Fig. \ref{Fig3}, we show the resulting topological band structure and density of states (D.O.S.) in the single particle picture. The considered unit cell with $p=7, q = 4$ lattice sites provides four bands as lattice $\rm{A}$ has $q = 4$ sites in its magnetic unit cell. 
We observe that the optimized lattice exhibits topological flat bands as shown in Fig. \ref{Fig3} (b) for nearest-neighbor hoppings. Fig. \ref{Fig3} (a) depicts the results if we do not optimize the positions of lattice ${\rm B}$ sites.
Notably, for the optimized lattice, the density of states is gapped and remains gapped if we take into account the long-range character of the dipolar interaction as shown in Fig. \ref{Fig3} (c).

In summary, Fig. \ref{Fig3} shows that we have obtained a flat band for a system with $p/q = 7/4$. However, we want to point out that flat topological bands can be achieved for arbitrary flux values by adjusting the ratio $p/q$ and optimizing the magnetic unit cell through our proposed method.

\section{Many-Body Physics}
In the single-particle picture, we have obtained nearly homogeneous flux and flat bands in the Rydberg lattices. Now, we extend our analysis to the many-body regime. To be specific, we aim to study a fractional Chern insulator (FCI) state, using the flux generated in the lattice through our method. The properties of our model in the single-particle regime combined with the strong interaction between particles provided by the hard-core constraint, make our approach a promising tool for realizing FCI states using Rydberg atoms.

We use exact diagonalization (ED) method to obtain the many-body ground state of our Hamiltonian provided by Eq.~(\ref{haml}). In our implementation of ED, the Hilbert space dimension is reduced due to particle number conservation. To obtain the FCI state, we consider finite lattices with $p/q = 7/4$ and a filling fraction $\nu = 1/2$. This corresponds to a lattice with $1/4$ flux per plaquette and one particle for every two flux. 
We consider lattice sizes from $4 \times 4$ to $8 \times 6$ with periodic boundary conditions and obtain the many-body ground state. Due to periodic boundary conditions, the topology of the system is a torus. The observed ground state is degenerate which is the signature of a FCI $\nu = 1/2$ state. To further support this interpretation of the ground state, we calculate the many-body Chern number for the ground state. For this, we use twisted periodic boundary conditions along $x$ and $y$ directions with twist angles $\theta_x$ and $\theta_y$
\begin{align}
    T_x(L_x)\psi(x,y) = e^{i\theta_x}\psi(x,y)\;, \nonumber \\
    T_y(L_y)\psi(x,y) = e^{i\theta_y}\psi(x,y)\;,
\end{align}
where $T_{i}(L_i)$ are operators which translate single-particle wavefunctions $\psi$ around the $i$th direction of the torus. The twist angles $\theta_x$ and $\theta_y$ correspond to additional fluxes which are threaded through the hole of the torus or through its interior. Thus, the many-body Hamiltonian and its eigenstates $\Psi(\theta_x,\theta_y)$ become functions of the twist angles. Within this parameter space, the non-abelian Berry connection and the many-body Chern number are defined as:
\begin{gather}
    \mathcal{A}_j^{\alpha,\beta}(\theta_x,\theta_y) = i\bra{\Psi^\alpha}\partial_j\ket{\Psi^\beta}\;, \\
    \mathcal{C} = \frac{1}{2\pi}\int_{0}^{2\pi}\int_{0}^{2\pi} \Tr{\partial_x\mathcal{A}_y^{\alpha,\beta} - \partial_y\mathcal{A}_x^{\alpha,\beta}} d\theta_x d\theta_y\;,
    \label{chern}
\end{gather}
where $\partial_j$ is the partial derivative with respect to $\theta_j$, and $\alpha,~\beta$ correspond to degenerate eigenstates. We calculate the many-body Chern number in addition to the ground state degeneracy to verify topological equivalence to Laughlin $\nu = 1/2$ states.

\begin{figure}
    \centering
    \includegraphics[width=\columnwidth]{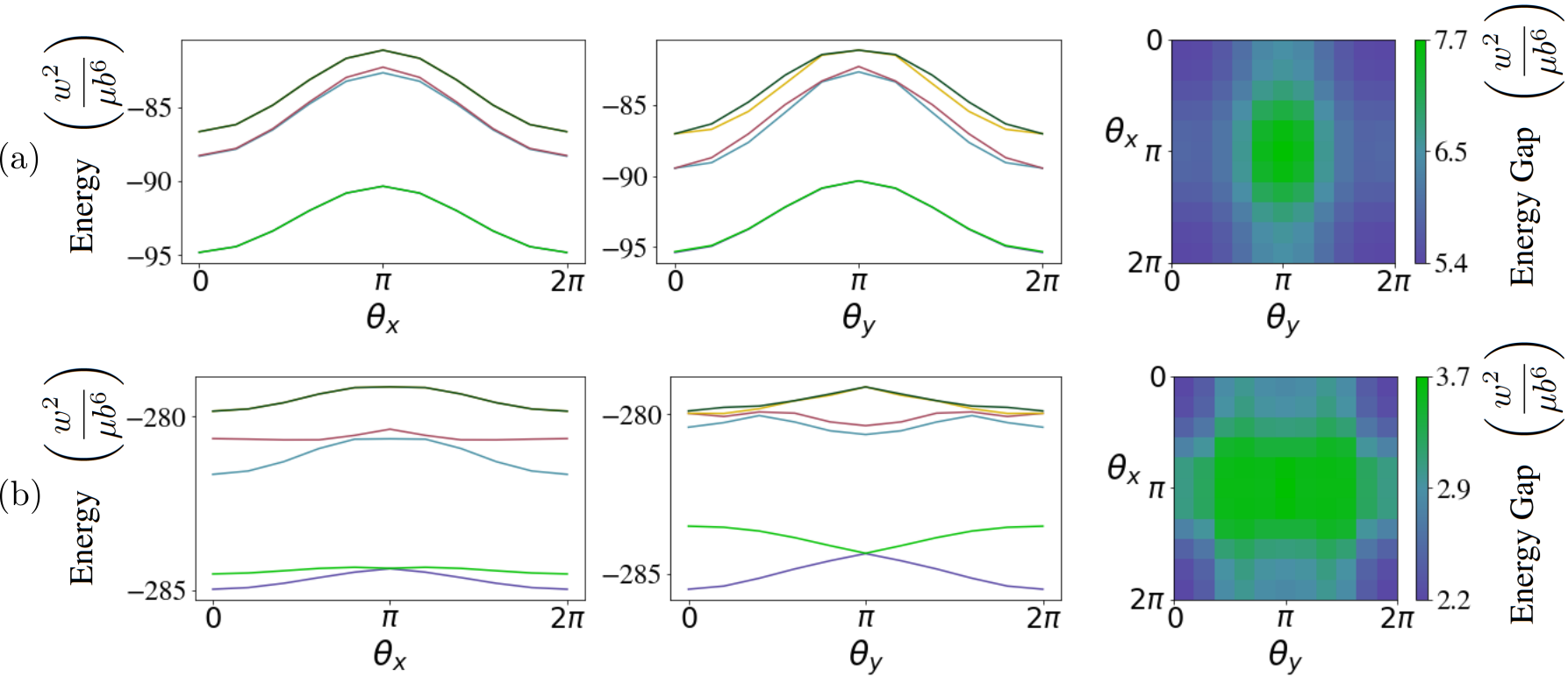}
    \caption{Results of the exact diagonalization for optimized $p/q=7/4$ lattices with half filling. In the left and middle plots, the lowest six eigenenergies are plotted against $\theta_x$ and $\theta_y$ while the other twist angle is held constant at $\pi$. In the right plots, the gap between the second and third lowest eigenenergies is shown across the whole twist angle space. (a) Lattice size of $4\times4$ with nearest-neighbor hopping. Here the degeneracy of the lowest two eigenenergies and their many-body Chern number $\mathcal{C}=1$ correspond with the Laughlin $\nu=1/2$ state. (b) Lattice size of $8\times6$ with nearest-neighbor hopping. The lowest two eigenstates remain degenerate and gapped to excitation, and the calculated many-body Chern number for the lowest two eigenstates remains $\mathcal{C}=1$.}
    \label{Fig4}
\end{figure}

The lowest six eigenenergies for nearest-neighbor hopping as a function of twist angles is shown in Fig. \ref{Fig4} (a) and (b) for lattices of size $4 \times 4$ and $8 \times 6$, respectively. Here, the results are computed across an $11\times11$ discretization of the twist angle space. We observe that the ground state is degenerate and it is gapped from the first excited state for all twist angles. The doubly-degenerate ground state manifold has a Chern number $\mathcal{C}=1$. This corresponds with the FCI $\nu=1/2$ state. The nearest-neighbor model also maps to the known models of FQH states considered in Ref. \cite{bai_18}. Therefore, we have established that our approach exhibits FCI state if we consider hopping up to nearest-neighbor.
To make a conclusive statement about the effect of long-range hoppings, larger systems would be required, which are not accessible via exact diagonalization. However, already the reduced flatness in the single-particle band structure, see Fig \ref{Fig3} (c), indicates that the topological state might be destabilized. On the other hand, it might be that more sophisticated and numerically challenging optimization strategies than those applied by us can find geometries that result in flatter bands. Thus, it is an open question whether, in the thermodynamic limit, the topological state survives in the presence of long-range hoppings.

\section{Conclusion}
We have proposed a method to generate arbitrary homogeneous flux and hopping in rectangular lattices of Rydberg atoms using dipolar exchange interactions and a background lattice of ancilla atoms.
Note that while we focused on Rydberg atoms, our proposal can be easily adapted to polar molecules, where similar level structures and interactions can be obtained.
We showed that within a magnetic unit cell, the flux is proportional to the ratio between the number of sites of the background lattice, $p$, and the number of sites of the rectangular lattice, $q$. We demonstrated that the flux is topologically robust and unaffected by changes in the atomic positions within the magnetic unit cell as long as hopping amplitudes do not vanish. To make the flux homogeneous, we made use of the freedom to change the atomic positions. With this homogeneous flux, we obtained topological flat bands in the single-particle regime. Further, in the many-body regime, we obtained a bosonic fractional Chern insulator state at $\nu = 1/2$ filling with nearest neighbor hopping. This many-body state has the same topological properties as of the Laughlin state $\nu = 1/2$ with many body Chern number $\mathcal{C} = 1$.
To fully understand the effect of long-range hoppings, larger system sizes must be explored. These were inaccessible by the exact diagonalization method applied by us and require numerical methods such as DMRG. Alternatively, to suppress long-range hoppings altogether, one can implement our model using Rydberg $S$ and $D$ states that are only dipole-coupled in second order, giving rise to exchange interactions that scale with $1/R^6$ where $R$ is the interatomic distance. Future work can be done to explore the entire parameter space of our model.\\

\acknowledgments{We gratefully acknowledge financial support by the Baden-W\"urttemberg Stiftung via Grant BWST ISF2019-017
under the program Internationale Spitzenforschung. This project has also received funding  from the French-German collaboration for
joint projects in NLE Sciences funded by the Deutsche Forschungsgemeinschaft (DFG) and the Agence National de la Recherche (ANR, project RYBOTIN). A.B. and T.L. acknowledge funding by the Agence Nationale de la Recherche (ANR-22-PETQ-0004 France 2030, project QuBitAF), Horizon Europe programme HORIZON-CL4-2022-QUANTUM-02-SGA via the project 101113690 (PASQuanS2.1), and the European Research Council (Advanced grant No. 101018511-ATARAXIA).}

\bibliography{ref}{}

\begin{thebibliography}{43}%
\makeatletter
\providecommand \@ifxundefined [1]{%
 \@ifx{#1\undefined}
}%
\providecommand \@ifnum [1]{%
 \ifnum #1\expandafter \@firstoftwo
 \else \expandafter \@secondoftwo
 \fi
}%
\providecommand \@ifx [1]{%
 \ifx #1\expandafter \@firstoftwo
 \else \expandafter \@secondoftwo
 \fi
}%
\providecommand \natexlab [1]{#1}%
\providecommand \enquote  [1]{``#1''}%
\providecommand \bibnamefont  [1]{#1}%
\providecommand \bibfnamefont [1]{#1}%
\providecommand \citenamefont [1]{#1}%
\providecommand \href@noop [0]{\@secondoftwo}%
\providecommand \href [0]{\begingroup \@sanitize@url \@href}%
\providecommand \@href[1]{\@@startlink{#1}\@@href}%
\providecommand \@@href[1]{\endgroup#1\@@endlink}%
\providecommand \@sanitize@url [0]{\catcode `\\12\catcode `\$12\catcode
  `\&12\catcode `\#12\catcode `\^12\catcode `\_12\catcode `\%12\relax}%
\providecommand \@@startlink[1]{}%
\providecommand \@@endlink[0]{}%
\providecommand \url  [0]{\begingroup\@sanitize@url \@url }%
\providecommand \@url [1]{\endgroup\@href {#1}{\urlprefix }}%
\providecommand \urlprefix  [0]{URL }%
\providecommand \Eprint [0]{\href }%
\providecommand \doibase [0]{https://doi.org/}%
\providecommand \selectlanguage [0]{\@gobble}%
\providecommand \bibinfo  [0]{\@secondoftwo}%
\providecommand \bibfield  [0]{\@secondoftwo}%
\providecommand \translation [1]{[#1]}%
\providecommand \BibitemOpen [0]{}%
\providecommand \bibitemStop [0]{}%
\providecommand \bibitemNoStop [0]{.\EOS\space}%
\providecommand \EOS [0]{\spacefactor3000\relax}%
\providecommand \BibitemShut  [1]{\csname bibitem#1\endcsname}%
\let\auto@bib@innerbib\@empty
\bibitem [{\citenamefont {Hofstadter}(1976)}]{hofstadter_76}%
  \BibitemOpen
  \bibfield  {author} {\bibinfo {author} {\bibfnamefont {D.~R.}\ \bibnamefont
  {Hofstadter}},\ }\bibfield  {title} {\bibinfo {title} {Energy levels and wave
  functions of {B}loch electrons in rational and irrational magnetic fields},\
  }\href {https://doi.org/10.1103/PhysRevB.14.2239} {\bibfield  {journal}
  {\bibinfo  {journal} {Phys. Rev. B}\ }\textbf {\bibinfo {volume} {14}},\
  \bibinfo {pages} {2239} (\bibinfo {year} {1976})}\BibitemShut {NoStop}%
\bibitem [{\citenamefont {Laughlin}(1983)}]{laughlin_83}%
  \BibitemOpen
  \bibfield  {author} {\bibinfo {author} {\bibfnamefont {R.~B.}\ \bibnamefont
  {Laughlin}},\ }\bibfield  {title} {\bibinfo {title} {Anomalous quantum {H}all
  effect: An incompressible quantum fluid with fractionally charged
  excitations},\ }\href {https://doi.org/10.1103/PhysRevLett.50.1395}
  {\bibfield  {journal} {\bibinfo  {journal} {Phys. Rev. Lett.}\ }\textbf
  {\bibinfo {volume} {50}},\ \bibinfo {pages} {1395} (\bibinfo {year}
  {1983})}\BibitemShut {NoStop}%
\bibitem [{\citenamefont {Thouless}\ \emph {et~al.}(1982)\citenamefont
  {Thouless}, \citenamefont {Kohmoto}, \citenamefont {Nightingale},\ and\
  \citenamefont {den Nijs}}]{Nijs_82}%
  \BibitemOpen
  \bibfield  {author} {\bibinfo {author} {\bibfnamefont {D.~J.}\ \bibnamefont
  {Thouless}}, \bibinfo {author} {\bibfnamefont {M.}~\bibnamefont {Kohmoto}},
  \bibinfo {author} {\bibfnamefont {M.~P.}\ \bibnamefont {Nightingale}},\ and\
  \bibinfo {author} {\bibfnamefont {M.}~\bibnamefont {den Nijs}},\ }\bibfield
  {title} {\bibinfo {title} {Quantized hall conductance in a two-dimensional
  periodic potential},\ }\href {https://doi.org/10.1103/PhysRevLett.49.405}
  {\bibfield  {journal} {\bibinfo  {journal} {Phys. Rev. Lett.}\ }\textbf
  {\bibinfo {volume} {49}},\ \bibinfo {pages} {405} (\bibinfo {year}
  {1982})}\BibitemShut {NoStop}%
\bibitem [{\citenamefont {Halperin}(1982)}]{Halperin_82}%
  \BibitemOpen
  \bibfield  {author} {\bibinfo {author} {\bibfnamefont {B.~I.}\ \bibnamefont
  {Halperin}},\ }\bibfield  {title} {\bibinfo {title} {Quantized hall
  conductance, current-carrying edge states, and the existence of extended
  states in a two-dimensional disordered potential},\ }\href
  {https://doi.org/10.1103/PhysRevB.25.2185} {\bibfield  {journal} {\bibinfo
  {journal} {Phys. Rev. B}\ }\textbf {\bibinfo {volume} {25}},\ \bibinfo
  {pages} {2185} (\bibinfo {year} {1982})}\BibitemShut {NoStop}%
\bibitem [{\citenamefont {Simon}(1983)}]{simon_83}%
  \BibitemOpen
  \bibfield  {author} {\bibinfo {author} {\bibfnamefont {B.}~\bibnamefont
  {Simon}},\ }\bibfield  {title} {\bibinfo {title} {Holonomy, the quantum
  adiabatic theorem, and berry's phase},\ }\href
  {https://doi.org/10.1103/PhysRevLett.51.2167} {\bibfield  {journal} {\bibinfo
   {journal} {Phys. Rev. Lett.}\ }\textbf {\bibinfo {volume} {51}},\ \bibinfo
  {pages} {2167} (\bibinfo {year} {1983})}\BibitemShut {NoStop}%
\bibitem [{\citenamefont {Kohmoto}(1985)}]{kohmoto_85}%
  \BibitemOpen
  \bibfield  {author} {\bibinfo {author} {\bibfnamefont {M.}~\bibnamefont
  {Kohmoto}},\ }\bibfield  {title} {\bibinfo {title} {Topological invariant and
  the quantization of the hall conductance},\ }\href
  {https://doi.org/https://doi.org/10.1016/0003-4916(85)90148-4} {\bibfield
  {journal} {\bibinfo  {journal} {Annals of Physics}\ }\textbf {\bibinfo
  {volume} {160}},\ \bibinfo {pages} {343} (\bibinfo {year}
  {1985})}\BibitemShut {NoStop}%
\bibitem [{\citenamefont {Hatsugai}(1993)}]{Hatsugai_93}%
  \BibitemOpen
  \bibfield  {author} {\bibinfo {author} {\bibfnamefont {Y.}~\bibnamefont
  {Hatsugai}},\ }\bibfield  {title} {\bibinfo {title} {Chern number and edge
  states in the integer quantum hall effect},\ }\href
  {https://doi.org/10.1103/PhysRevLett.71.3697} {\bibfield  {journal} {\bibinfo
   {journal} {Phys. Rev. Lett.}\ }\textbf {\bibinfo {volume} {71}},\ \bibinfo
  {pages} {3697} (\bibinfo {year} {1993})}\BibitemShut {NoStop}%
\bibitem [{\citenamefont {Klitzing}\ \emph {et~al.}(1980)\citenamefont
  {Klitzing}, \citenamefont {Dorda},\ and\ \citenamefont
  {Pepper}}]{klitzing_80}%
  \BibitemOpen
  \bibfield  {author} {\bibinfo {author} {\bibfnamefont {K.~v.}\ \bibnamefont
  {Klitzing}}, \bibinfo {author} {\bibfnamefont {G.}~\bibnamefont {Dorda}},\
  and\ \bibinfo {author} {\bibfnamefont {M.}~\bibnamefont {Pepper}},\
  }\bibfield  {title} {\bibinfo {title} {New method for high-accuracy
  determination of the fine-structure constant based on quantized hall
  resistance},\ }\href {https://doi.org/10.1103/PhysRevLett.45.494} {\bibfield
  {journal} {\bibinfo  {journal} {Phys. Rev. Lett.}\ }\textbf {\bibinfo
  {volume} {45}},\ \bibinfo {pages} {494} (\bibinfo {year} {1980})}\BibitemShut
  {NoStop}%
\bibitem [{\citenamefont {Tsui}\ \emph {et~al.}(1982)\citenamefont {Tsui},
  \citenamefont {Stormer},\ and\ \citenamefont {Gossard}}]{stormer_82}%
  \BibitemOpen
  \bibfield  {author} {\bibinfo {author} {\bibfnamefont {D.~C.}\ \bibnamefont
  {Tsui}}, \bibinfo {author} {\bibfnamefont {H.~L.}\ \bibnamefont {Stormer}},\
  and\ \bibinfo {author} {\bibfnamefont {A.~C.}\ \bibnamefont {Gossard}},\
  }\bibfield  {title} {\bibinfo {title} {Two-dimensional magnetotransport in
  the extreme quantum limit},\ }\href
  {https://doi.org/10.1103/PhysRevLett.48.1559} {\bibfield  {journal} {\bibinfo
   {journal} {Phys. Rev. Lett.}\ }\textbf {\bibinfo {volume} {48}},\ \bibinfo
  {pages} {1559} (\bibinfo {year} {1982})}\BibitemShut {NoStop}%
\bibitem [{\citenamefont {Regnault}\ and\ \citenamefont
  {Bernevig}(2011)}]{regnault_11}%
  \BibitemOpen
  \bibfield  {author} {\bibinfo {author} {\bibfnamefont {N.}~\bibnamefont
  {Regnault}}\ and\ \bibinfo {author} {\bibfnamefont {B.~A.}\ \bibnamefont
  {Bernevig}},\ }\bibfield  {title} {\bibinfo {title} {Fractional chern
  insulator},\ }\href {https://doi.org/10.1103/PhysRevX.1.021014} {\bibfield
  {journal} {\bibinfo  {journal} {Phys. Rev. X}\ }\textbf {\bibinfo {volume}
  {1}},\ \bibinfo {pages} {021014} (\bibinfo {year} {2011})}\BibitemShut
  {NoStop}%
\bibitem [{\citenamefont {Bloch}\ \emph {et~al.}(2012)\citenamefont {Bloch},
  \citenamefont {Dalibard},\ and\ \citenamefont {Nascimbène}}]{bloch_12}%
  \BibitemOpen
  \bibfield  {author} {\bibinfo {author} {\bibfnamefont {I.}~\bibnamefont
  {Bloch}}, \bibinfo {author} {\bibfnamefont {J.}~\bibnamefont {Dalibard}},\
  and\ \bibinfo {author} {\bibfnamefont {S.}~\bibnamefont {Nascimbène}},\
  }\bibfield  {title} {\bibinfo {title} {Quantum simulations with ultracold
  quantum gases},\ }\href {https://doi.org/10.1038/nphys2259} {\bibfield
  {journal} {\bibinfo  {journal} {Nature Physics}\ }\textbf {\bibinfo {volume}
  {8}},\ \bibinfo {pages} {267} (\bibinfo {year} {2012})}\BibitemShut {NoStop}%
\bibitem [{\citenamefont {Gross}\ and\ \citenamefont {Bloch}(2017)}]{gross_17}%
  \BibitemOpen
  \bibfield  {author} {\bibinfo {author} {\bibfnamefont {C.}~\bibnamefont
  {Gross}}\ and\ \bibinfo {author} {\bibfnamefont {I.}~\bibnamefont {Bloch}},\
  }\bibfield  {title} {\bibinfo {title} {Quantum simulations with ultracold
  atoms in optical lattices},\ }\href {https://doi.org/10.1126/science.aal3837}
  {\bibfield  {journal} {\bibinfo  {journal} {Science}\ }\textbf {\bibinfo
  {volume} {357}},\ \bibinfo {pages} {995} (\bibinfo {year}
  {2017})}\BibitemShut {NoStop}%
\bibitem [{\citenamefont {Sch\"afer}\ \emph {et~al.}(2020)\citenamefont
  {Sch\"afer}, \citenamefont {Fukuhara}, \citenamefont {Sugawa}, \citenamefont
  {Takasu},\ and\ \citenamefont {Takahashi}}]{florian_20}%
  \BibitemOpen
  \bibfield  {author} {\bibinfo {author} {\bibfnamefont {F.}~\bibnamefont
  {Sch\"afer}}, \bibinfo {author} {\bibfnamefont {T.}~\bibnamefont {Fukuhara}},
  \bibinfo {author} {\bibfnamefont {S.}~\bibnamefont {Sugawa}}, \bibinfo
  {author} {\bibfnamefont {Y.}~\bibnamefont {Takasu}},\ and\ \bibinfo {author}
  {\bibfnamefont {Y.}~\bibnamefont {Takahashi}},\ }\bibfield  {title} {\bibinfo
  {title} {Tools for quantum simulation with ultracold atoms in optical
  lattices},\ }\href {https://doi.org/10.1038/s42254-020-0195-3} {\bibfield
  {journal} {\bibinfo  {journal} {Nature Reviews Physics}\ }\textbf {\bibinfo
  {volume} {2}},\ \bibinfo {pages} {411} (\bibinfo {year} {2020})}\BibitemShut
  {NoStop}%
\bibitem [{\citenamefont {Jaksch}\ \emph {et~al.}(1998)\citenamefont {Jaksch},
  \citenamefont {Bruder}, \citenamefont {Cirac}, \citenamefont {Gardiner},\
  and\ \citenamefont {Zoller}}]{jaksch_98}%
  \BibitemOpen
  \bibfield  {author} {\bibinfo {author} {\bibfnamefont {D.}~\bibnamefont
  {Jaksch}}, \bibinfo {author} {\bibfnamefont {C.}~\bibnamefont {Bruder}},
  \bibinfo {author} {\bibfnamefont {J.~I.}\ \bibnamefont {Cirac}}, \bibinfo
  {author} {\bibfnamefont {C.~W.}\ \bibnamefont {Gardiner}},\ and\ \bibinfo
  {author} {\bibfnamefont {P.}~\bibnamefont {Zoller}},\ }\bibfield  {title}
  {\bibinfo {title} {Cold bosonic atoms in optical lattices},\ }\href
  {https://doi.org/10.1103/PhysRevLett.81.3108} {\bibfield  {journal} {\bibinfo
   {journal} {Phys. Rev. Lett.}\ }\textbf {\bibinfo {volume} {81}},\ \bibinfo
  {pages} {3108} (\bibinfo {year} {1998})}\BibitemShut {NoStop}%
\bibitem [{\citenamefont {Greiner}\ \emph {et~al.}(2001)\citenamefont
  {Greiner}, \citenamefont {Bloch}, \citenamefont {Mandel}, \citenamefont
  {H\"ansch},\ and\ \citenamefont {Esslinger}}]{greiner_01}%
  \BibitemOpen
  \bibfield  {author} {\bibinfo {author} {\bibfnamefont {M.}~\bibnamefont
  {Greiner}}, \bibinfo {author} {\bibfnamefont {I.}~\bibnamefont {Bloch}},
  \bibinfo {author} {\bibfnamefont {O.}~\bibnamefont {Mandel}}, \bibinfo
  {author} {\bibfnamefont {T.~W.}\ \bibnamefont {H\"ansch}},\ and\ \bibinfo
  {author} {\bibfnamefont {T.}~\bibnamefont {Esslinger}},\ }\bibfield  {title}
  {\bibinfo {title} {Exploring phase coherence in a 2{D} lattice of
  {B}ose-{E}instein condensates},\ }\href
  {https://doi.org/10.1103/PhysRevLett.87.160405} {\bibfield  {journal}
  {\bibinfo  {journal} {Phys. Rev. Lett.}\ }\textbf {\bibinfo {volume} {87}},\
  \bibinfo {pages} {160405} (\bibinfo {year} {2001})}\BibitemShut {NoStop}%
\bibitem [{\citenamefont {Greiner}\ \emph {et~al.}(2002)\citenamefont
  {Greiner}, \citenamefont {Mandel}, \citenamefont {Esslinger}, \citenamefont
  {H\"ansch},\ and\ \citenamefont {Bloch}}]{greiner_02}%
  \BibitemOpen
  \bibfield  {author} {\bibinfo {author} {\bibfnamefont {M.}~\bibnamefont
  {Greiner}}, \bibinfo {author} {\bibfnamefont {O.}~\bibnamefont {Mandel}},
  \bibinfo {author} {\bibfnamefont {T.}~\bibnamefont {Esslinger}}, \bibinfo
  {author} {\bibfnamefont {T.~W.}\ \bibnamefont {H\"ansch}},\ and\ \bibinfo
  {author} {\bibfnamefont {I.}~\bibnamefont {Bloch}},\ }\bibfield  {title}
  {\bibinfo {title} {Quantum phase transition from a superfluid to a {M}ott
  insulator in a gas of ultracold atoms},\ }\href
  {https://doi.org/10.1038/415039a} {\bibfield  {journal} {\bibinfo  {journal}
  {Nature (London)}\ }\textbf {\bibinfo {volume} {415}},\ \bibinfo {pages} {39}
  (\bibinfo {year} {2002})}\BibitemShut {NoStop}%
\bibitem [{\citenamefont {Jaksch}\ and\ \citenamefont
  {Zoller}(2003)}]{jaksch_03}%
  \BibitemOpen
  \bibfield  {author} {\bibinfo {author} {\bibfnamefont {D.}~\bibnamefont
  {Jaksch}}\ and\ \bibinfo {author} {\bibfnamefont {P.}~\bibnamefont
  {Zoller}},\ }\bibfield  {title} {\bibinfo {title} {Creation of effective
  magnetic fields in optical lattices: the {H}ofstadter butterfly for cold
  neutral atoms},\ }\href
  {https://doi.org/https://doi.org/10.1088/1367-2630/5/1/356} {\bibfield
  {journal} {\bibinfo  {journal} {New J. Phys.}\ }\textbf {\bibinfo {volume}
  {5}},\ \bibinfo {pages} {56} (\bibinfo {year} {2003})}\BibitemShut {NoStop}%
\bibitem [{\citenamefont {Dalibard}\ \emph {et~al.}(2011)\citenamefont
  {Dalibard}, \citenamefont {Gerbier}, \citenamefont
  {Juzeli\ifmmode~\bar{u}\else \={u}\fi{}nas},\ and\ \citenamefont
  {\"Ohberg}}]{delibard_11}%
  \BibitemOpen
  \bibfield  {author} {\bibinfo {author} {\bibfnamefont {J.}~\bibnamefont
  {Dalibard}}, \bibinfo {author} {\bibfnamefont {F.}~\bibnamefont {Gerbier}},
  \bibinfo {author} {\bibfnamefont {G.}~\bibnamefont
  {Juzeli\ifmmode~\bar{u}\else \={u}\fi{}nas}},\ and\ \bibinfo {author}
  {\bibfnamefont {P.}~\bibnamefont {\"Ohberg}},\ }\bibfield  {title} {\bibinfo
  {title} {Colloquium: Artificial gauge potentials for neutral atoms},\ }\href
  {https://doi.org/10.1103/RevModPhys.83.1523} {\bibfield  {journal} {\bibinfo
  {journal} {Rev. Mod. Phys.}\ }\textbf {\bibinfo {volume} {83}},\ \bibinfo
  {pages} {1523} (\bibinfo {year} {2011})}\BibitemShut {NoStop}%
\bibitem [{\citenamefont {Lin}\ \emph {et~al.}(2009{\natexlab{a}})\citenamefont
  {Lin}, \citenamefont {Compton}, \citenamefont {Jimenez-Garcia}, \citenamefont
  {Porto},\ and\ \citenamefont {Spielman}}]{lin_09b}%
  \BibitemOpen
  \bibfield  {author} {\bibinfo {author} {\bibfnamefont {Y.-J.}\ \bibnamefont
  {Lin}}, \bibinfo {author} {\bibfnamefont {R.~L.}\ \bibnamefont {Compton}},
  \bibinfo {author} {\bibfnamefont {K.}~\bibnamefont {Jimenez-Garcia}},
  \bibinfo {author} {\bibfnamefont {J.~V.}\ \bibnamefont {Porto}},\ and\
  \bibinfo {author} {\bibfnamefont {I.~B.}\ \bibnamefont {Spielman}},\
  }\bibfield  {title} {\bibinfo {title} {Synthetic magnetic fields for
  ultracold neutral atoms},\ }\href {https://doi.org/10.1038/nature08609}
  {\bibfield  {journal} {\bibinfo  {journal} {Nature (London)}\ }\textbf
  {\bibinfo {volume} {462}},\ \bibinfo {pages} {628} (\bibinfo {year}
  {2009}{\natexlab{a}})}\BibitemShut {NoStop}%
\bibitem [{\citenamefont {Lin}\ \emph {et~al.}(2009{\natexlab{b}})\citenamefont
  {Lin}, \citenamefont {Compton}, \citenamefont {Perry}, \citenamefont
  {Phillips}, \citenamefont {Porto},\ and\ \citenamefont {Spielman}}]{lin_09a}%
  \BibitemOpen
  \bibfield  {author} {\bibinfo {author} {\bibfnamefont {Y.-J.}\ \bibnamefont
  {Lin}}, \bibinfo {author} {\bibfnamefont {R.~L.}\ \bibnamefont {Compton}},
  \bibinfo {author} {\bibfnamefont {A.~R.}\ \bibnamefont {Perry}}, \bibinfo
  {author} {\bibfnamefont {W.~D.}\ \bibnamefont {Phillips}}, \bibinfo {author}
  {\bibfnamefont {J.~V.}\ \bibnamefont {Porto}},\ and\ \bibinfo {author}
  {\bibfnamefont {I.~B.}\ \bibnamefont {Spielman}},\ }\bibfield  {title}
  {\bibinfo {title} {{B}ose-{E}instein condensate in a uniform light-induced
  vector potential},\ }\href {https://doi.org/10.1103/PhysRevLett.102.130401}
  {\bibfield  {journal} {\bibinfo  {journal} {Phys. Rev. Lett.}\ }\textbf
  {\bibinfo {volume} {102}},\ \bibinfo {pages} {130401} (\bibinfo {year}
  {2009}{\natexlab{b}})}\BibitemShut {NoStop}%
\bibitem [{\citenamefont {Lin}\ \emph {et~al.}(2011)\citenamefont {Lin},
  \citenamefont {Compton}, \citenamefont {Jimenez-Garcia}, \citenamefont
  {Phillips}, \citenamefont {Porto},\ and\ \citenamefont {Spielman}}]{lin_11}%
  \BibitemOpen
  \bibfield  {author} {\bibinfo {author} {\bibfnamefont {Y.-J.}\ \bibnamefont
  {Lin}}, \bibinfo {author} {\bibfnamefont {R.~L.}\ \bibnamefont {Compton}},
  \bibinfo {author} {\bibfnamefont {K.}~\bibnamefont {Jimenez-Garcia}},
  \bibinfo {author} {\bibfnamefont {W.~D.}\ \bibnamefont {Phillips}}, \bibinfo
  {author} {\bibfnamefont {J.~V.}\ \bibnamefont {Porto}},\ and\ \bibinfo
  {author} {\bibfnamefont {I.~B.}\ \bibnamefont {Spielman}},\ }\bibfield
  {title} {\bibinfo {title} {A synthetic electric force acting on neutral
  atoms},\ }\href {https://doi.org/10.1038/nphys1954} {\bibfield  {journal}
  {\bibinfo  {journal} {Nat. Phys.}\ }\textbf {\bibinfo {volume} {7}},\
  \bibinfo {pages} {531} (\bibinfo {year} {2011})}\BibitemShut {NoStop}%
\bibitem [{\citenamefont {Aidelsburger}\ \emph {et~al.}(2011)\citenamefont
  {Aidelsburger}, \citenamefont {Atala}, \citenamefont {Nascimb\`ene},
  \citenamefont {Trotzky}, \citenamefont {Chen},\ and\ \citenamefont
  {Bloch}}]{aidelsburger_11}%
  \BibitemOpen
  \bibfield  {author} {\bibinfo {author} {\bibfnamefont {M.}~\bibnamefont
  {Aidelsburger}}, \bibinfo {author} {\bibfnamefont {M.}~\bibnamefont {Atala}},
  \bibinfo {author} {\bibfnamefont {S.}~\bibnamefont {Nascimb\`ene}}, \bibinfo
  {author} {\bibfnamefont {S.}~\bibnamefont {Trotzky}}, \bibinfo {author}
  {\bibfnamefont {Y.-A.}\ \bibnamefont {Chen}},\ and\ \bibinfo {author}
  {\bibfnamefont {I.}~\bibnamefont {Bloch}},\ }\bibfield  {title} {\bibinfo
  {title} {Experimental realization of strong effective magnetic fields in an
  optical lattice},\ }\href {https://doi.org/10.1103/PhysRevLett.107.255301}
  {\bibfield  {journal} {\bibinfo  {journal} {Phys. Rev. Lett.}\ }\textbf
  {\bibinfo {volume} {107}},\ \bibinfo {pages} {255301} (\bibinfo {year}
  {2011})}\BibitemShut {NoStop}%
\bibitem [{\citenamefont {Miyake}\ \emph {et~al.}(2013)\citenamefont {Miyake},
  \citenamefont {Siviloglou}, \citenamefont {Kennedy}, \citenamefont {Burton},\
  and\ \citenamefont {Ketterle}}]{miyake_13}%
  \BibitemOpen
  \bibfield  {author} {\bibinfo {author} {\bibfnamefont {H.}~\bibnamefont
  {Miyake}}, \bibinfo {author} {\bibfnamefont {G.~A.}\ \bibnamefont
  {Siviloglou}}, \bibinfo {author} {\bibfnamefont {C.~J.}\ \bibnamefont
  {Kennedy}}, \bibinfo {author} {\bibfnamefont {W.~C.}\ \bibnamefont
  {Burton}},\ and\ \bibinfo {author} {\bibfnamefont {W.}~\bibnamefont
  {Ketterle}},\ }\bibfield  {title} {\bibinfo {title} {Realizing the {H}arper
  {H}amiltonian with laser-assisted tunneling in optical lattices},\ }\href
  {https://doi.org/10.1103/PhysRevLett.111.185302} {\bibfield  {journal}
  {\bibinfo  {journal} {Phys. Rev. Lett.}\ }\textbf {\bibinfo {volume} {111}},\
  \bibinfo {pages} {185302} (\bibinfo {year} {2013})}\BibitemShut {NoStop}%
\bibitem [{\citenamefont {Williams}\ \emph {et~al.}(2010)\citenamefont
  {Williams}, \citenamefont {Al-Assam},\ and\ \citenamefont
  {Foot}}]{williams_10}%
  \BibitemOpen
  \bibfield  {author} {\bibinfo {author} {\bibfnamefont {R.~A.}\ \bibnamefont
  {Williams}}, \bibinfo {author} {\bibfnamefont {S.}~\bibnamefont {Al-Assam}},\
  and\ \bibinfo {author} {\bibfnamefont {C.~J.}\ \bibnamefont {Foot}},\
  }\bibfield  {title} {\bibinfo {title} {Observation of vortex nucleation in a
  rotating two-dimensional lattice of bose-einstein condensates},\ }\href
  {https://doi.org/10.1103/PhysRevLett.104.050404} {\bibfield  {journal}
  {\bibinfo  {journal} {Phys. Rev. Lett.}\ }\textbf {\bibinfo {volume} {104}},\
  \bibinfo {pages} {050404} (\bibinfo {year} {2010})}\BibitemShut {NoStop}%
\bibitem [{\citenamefont {Jim\'enez-Garc\'{\i}a}\ \emph
  {et~al.}(2012)\citenamefont {Jim\'enez-Garc\'{\i}a}, \citenamefont {LeBlanc},
  \citenamefont {Williams}, \citenamefont {Beeler}, \citenamefont {Perry},\
  and\ \citenamefont {Spielman}}]{jimenez_12}%
  \BibitemOpen
  \bibfield  {author} {\bibinfo {author} {\bibfnamefont {K.}~\bibnamefont
  {Jim\'enez-Garc\'{\i}a}}, \bibinfo {author} {\bibfnamefont {L.~J.}\
  \bibnamefont {LeBlanc}}, \bibinfo {author} {\bibfnamefont {R.~A.}\
  \bibnamefont {Williams}}, \bibinfo {author} {\bibfnamefont {M.~C.}\
  \bibnamefont {Beeler}}, \bibinfo {author} {\bibfnamefont {A.~R.}\
  \bibnamefont {Perry}},\ and\ \bibinfo {author} {\bibfnamefont {I.~B.}\
  \bibnamefont {Spielman}},\ }\bibfield  {title} {\bibinfo {title} {Peierls
  substitution in an engineered lattice potential},\ }\href
  {https://doi.org/10.1103/PhysRevLett.108.225303} {\bibfield  {journal}
  {\bibinfo  {journal} {Phys. Rev. Lett.}\ }\textbf {\bibinfo {volume} {108}},\
  \bibinfo {pages} {225303} (\bibinfo {year} {2012})}\BibitemShut {NoStop}%
\bibitem [{\citenamefont {Struck}\ \emph {et~al.}(2012)\citenamefont {Struck},
  \citenamefont {\"Olschl\"ager}, \citenamefont {Weinberg}, \citenamefont
  {Hauke}, \citenamefont {Simonet}, \citenamefont {Eckardt}, \citenamefont
  {Lewenstein}, \citenamefont {Sengstock},\ and\ \citenamefont
  {Windpassinger}}]{struck_12}%
  \BibitemOpen
  \bibfield  {author} {\bibinfo {author} {\bibfnamefont {J.}~\bibnamefont
  {Struck}}, \bibinfo {author} {\bibfnamefont {C.}~\bibnamefont
  {\"Olschl\"ager}}, \bibinfo {author} {\bibfnamefont {M.}~\bibnamefont
  {Weinberg}}, \bibinfo {author} {\bibfnamefont {P.}~\bibnamefont {Hauke}},
  \bibinfo {author} {\bibfnamefont {J.}~\bibnamefont {Simonet}}, \bibinfo
  {author} {\bibfnamefont {A.}~\bibnamefont {Eckardt}}, \bibinfo {author}
  {\bibfnamefont {M.}~\bibnamefont {Lewenstein}}, \bibinfo {author}
  {\bibfnamefont {K.}~\bibnamefont {Sengstock}},\ and\ \bibinfo {author}
  {\bibfnamefont {P.}~\bibnamefont {Windpassinger}},\ }\bibfield  {title}
  {\bibinfo {title} {Tunable gauge potential for neutral and spinless particles
  in driven optical lattices},\ }\href
  {https://doi.org/10.1103/PhysRevLett.108.225304} {\bibfield  {journal}
  {\bibinfo  {journal} {Phys. Rev. Lett.}\ }\textbf {\bibinfo {volume} {108}},\
  \bibinfo {pages} {225304} (\bibinfo {year} {2012})}\BibitemShut {NoStop}%
\bibitem [{\citenamefont {Price}\ \emph {et~al.}(2017)\citenamefont {Price},
  \citenamefont {Ozawa},\ and\ \citenamefont {Goldman}}]{price_17}%
  \BibitemOpen
  \bibfield  {author} {\bibinfo {author} {\bibfnamefont {H.~M.}\ \bibnamefont
  {Price}}, \bibinfo {author} {\bibfnamefont {T.}~\bibnamefont {Ozawa}},\ and\
  \bibinfo {author} {\bibfnamefont {N.}~\bibnamefont {Goldman}},\ }\bibfield
  {title} {\bibinfo {title} {Synthetic dimensions for cold atoms from shaking a
  harmonic trap},\ }\href {https://doi.org/10.1103/PhysRevA.95.023607}
  {\bibfield  {journal} {\bibinfo  {journal} {Phys. Rev. A}\ }\textbf {\bibinfo
  {volume} {95}},\ \bibinfo {pages} {023607} (\bibinfo {year}
  {2017})}\BibitemShut {NoStop}%
\bibitem [{\citenamefont {de~L\'es\'eleuc}\ \emph {et~al.}(2018)\citenamefont
  {de~L\'es\'eleuc}, \citenamefont {Weber}, \citenamefont {Lienhard},
  \citenamefont {Barredo}, \citenamefont {B\"uchler}, \citenamefont {Lahaye},\
  and\ \citenamefont {Browaeys}}]{deleleuc_18}%
  \BibitemOpen
  \bibfield  {author} {\bibinfo {author} {\bibfnamefont {S.}~\bibnamefont
  {de~L\'es\'eleuc}}, \bibinfo {author} {\bibfnamefont {S.}~\bibnamefont
  {Weber}}, \bibinfo {author} {\bibfnamefont {V.}~\bibnamefont {Lienhard}},
  \bibinfo {author} {\bibfnamefont {D.}~\bibnamefont {Barredo}}, \bibinfo
  {author} {\bibfnamefont {H.~P.}\ \bibnamefont {B\"uchler}}, \bibinfo {author}
  {\bibfnamefont {T.}~\bibnamefont {Lahaye}},\ and\ \bibinfo {author}
  {\bibfnamefont {A.}~\bibnamefont {Browaeys}},\ }\bibfield  {title} {\bibinfo
  {title} {Accurate mapping of multilevel rydberg atoms on interacting
  spin-$1/2$ particles for the quantum simulation of ising models},\ }\href
  {https://doi.org/10.1103/PhysRevLett.120.113602} {\bibfield  {journal}
  {\bibinfo  {journal} {Phys. Rev. Lett.}\ }\textbf {\bibinfo {volume} {120}},\
  \bibinfo {pages} {113602} (\bibinfo {year} {2018})}\BibitemShut {NoStop}%
\bibitem [{\citenamefont {Saffman}\ \emph {et~al.}(2010)\citenamefont
  {Saffman}, \citenamefont {Walker},\ and\ \citenamefont
  {M\o{}lmer}}]{saffman_10}%
  \BibitemOpen
  \bibfield  {author} {\bibinfo {author} {\bibfnamefont {M.}~\bibnamefont
  {Saffman}}, \bibinfo {author} {\bibfnamefont {T.~G.}\ \bibnamefont
  {Walker}},\ and\ \bibinfo {author} {\bibfnamefont {K.}~\bibnamefont
  {M\o{}lmer}},\ }\bibfield  {title} {\bibinfo {title} {Quantum information
  with rydberg atoms},\ }\href {https://doi.org/10.1103/RevModPhys.82.2313}
  {\bibfield  {journal} {\bibinfo  {journal} {Rev. Mod. Phys.}\ }\textbf
  {\bibinfo {volume} {82}},\ \bibinfo {pages} {2313} (\bibinfo {year}
  {2010})}\BibitemShut {NoStop}%
\bibitem [{\citenamefont {Weimer}\ \emph {et~al.}(2010)\citenamefont {Weimer},
  \citenamefont {M\"uller}, \citenamefont {Lesanovsky}, \citenamefont
  {Zoller},\ and\ \citenamefont {B\"uchler}}]{weimer_10}%
  \BibitemOpen
  \bibfield  {author} {\bibinfo {author} {\bibfnamefont {H.}~\bibnamefont
  {Weimer}}, \bibinfo {author} {\bibfnamefont {M.}~\bibnamefont {M\"uller}},
  \bibinfo {author} {\bibfnamefont {I.}~\bibnamefont {Lesanovsky}}, \bibinfo
  {author} {\bibfnamefont {P.}~\bibnamefont {Zoller}},\ and\ \bibinfo {author}
  {\bibfnamefont {H.~P.}\ \bibnamefont {B\"uchler}},\ }\bibfield  {title}
  {\bibinfo {title} {A rydberg quantum simulator},\ }\href
  {https://doi.org/10.1038/nphys1614} {\bibfield  {journal} {\bibinfo
  {journal} {Nature Physics}\ }\textbf {\bibinfo {volume} {6}},\ \bibinfo
  {pages} {382} (\bibinfo {year} {2010})}\BibitemShut {NoStop}%
\bibitem [{\citenamefont {Peierls}(1933)}]{peierls_33}%
  \BibitemOpen
  \bibfield  {author} {\bibinfo {author} {\bibfnamefont {R.~E.}\ \bibnamefont
  {Peierls}},\ }\bibfield  {title} {\bibinfo {title} {On the theory of
  diamagnetism of conduction electrons},\ }\href
  {https://doi.org/10.1007/BF01342591} {\bibfield  {journal} {\bibinfo
  {journal} {Z. Phys.}\ }\textbf {\bibinfo {volume} {80}},\ \bibinfo {pages}
  {763} (\bibinfo {year} {1933})}\BibitemShut {NoStop}%
\bibitem [{\citenamefont {Lienhard}\ \emph {et~al.}(2020)\citenamefont
  {Lienhard}, \citenamefont {Scholl}, \citenamefont {Weber}, \citenamefont
  {Barredo}, \citenamefont {de~L\'es\'eleuc}, \citenamefont {Bai},
  \citenamefont {Lang}, \citenamefont {Fleischhauer}, \citenamefont
  {B\"uchler}, \citenamefont {Lahaye},\ and\ \citenamefont
  {Browaeys}}]{lienhard_20}%
  \BibitemOpen
  \bibfield  {author} {\bibinfo {author} {\bibfnamefont {V.}~\bibnamefont
  {Lienhard}}, \bibinfo {author} {\bibfnamefont {P.}~\bibnamefont {Scholl}},
  \bibinfo {author} {\bibfnamefont {S.}~\bibnamefont {Weber}}, \bibinfo
  {author} {\bibfnamefont {D.}~\bibnamefont {Barredo}}, \bibinfo {author}
  {\bibfnamefont {S.}~\bibnamefont {de~L\'es\'eleuc}}, \bibinfo {author}
  {\bibfnamefont {R.}~\bibnamefont {Bai}}, \bibinfo {author} {\bibfnamefont
  {N.}~\bibnamefont {Lang}}, \bibinfo {author} {\bibfnamefont {M.}~\bibnamefont
  {Fleischhauer}}, \bibinfo {author} {\bibfnamefont {H.~P.}\ \bibnamefont
  {B\"uchler}}, \bibinfo {author} {\bibfnamefont {T.}~\bibnamefont {Lahaye}},\
  and\ \bibinfo {author} {\bibfnamefont {A.}~\bibnamefont {Browaeys}},\
  }\bibfield  {title} {\bibinfo {title} {Realization of a density-dependent
  peierls phase in a synthetic, spin-orbit coupled rydberg system},\ }\href
  {https://doi.org/10.1103/PhysRevX.10.021031} {\bibfield  {journal} {\bibinfo
  {journal} {Phys. Rev. X}\ }\textbf {\bibinfo {volume} {10}},\ \bibinfo
  {pages} {021031} (\bibinfo {year} {2020})}\BibitemShut {NoStop}%
\bibitem [{\citenamefont {Peter}\ \emph {et~al.}(2015)\citenamefont {Peter},
  \citenamefont {Yao}, \citenamefont {Lang}, \citenamefont {Huber},
  \citenamefont {Lukin},\ and\ \citenamefont {B\"uchler}}]{peter_15}%
  \BibitemOpen
  \bibfield  {author} {\bibinfo {author} {\bibfnamefont {D.}~\bibnamefont
  {Peter}}, \bibinfo {author} {\bibfnamefont {N.~Y.}\ \bibnamefont {Yao}},
  \bibinfo {author} {\bibfnamefont {N.}~\bibnamefont {Lang}}, \bibinfo {author}
  {\bibfnamefont {S.~D.}\ \bibnamefont {Huber}}, \bibinfo {author}
  {\bibfnamefont {M.~D.}\ \bibnamefont {Lukin}},\ and\ \bibinfo {author}
  {\bibfnamefont {H.~P.}\ \bibnamefont {B\"uchler}},\ }\bibfield  {title}
  {\bibinfo {title} {Topological bands with a chern number $c=2$ by dipolar
  exchange interactions},\ }\href {https://doi.org/10.1103/PhysRevA.91.053617}
  {\bibfield  {journal} {\bibinfo  {journal} {Phys. Rev. A}\ }\textbf {\bibinfo
  {volume} {91}},\ \bibinfo {pages} {053617} (\bibinfo {year}
  {2015})}\BibitemShut {NoStop}%
\bibitem [{\citenamefont {de~L{\'e}s{\'e}leuc}\ \emph
  {et~al.}(2019)\citenamefont {de~L{\'e}s{\'e}leuc}, \citenamefont {Lienhard},
  \citenamefont {Scholl}, \citenamefont {Barredo}, \citenamefont {Weber},
  \citenamefont {Lang}, \citenamefont {B{\"u}chler}, \citenamefont {Lahaye},\
  and\ \citenamefont {Browaeys}}]{deleleuc_19}%
  \BibitemOpen
  \bibfield  {author} {\bibinfo {author} {\bibfnamefont {S.}~\bibnamefont
  {de~L{\'e}s{\'e}leuc}}, \bibinfo {author} {\bibfnamefont {V.}~\bibnamefont
  {Lienhard}}, \bibinfo {author} {\bibfnamefont {P.}~\bibnamefont {Scholl}},
  \bibinfo {author} {\bibfnamefont {D.}~\bibnamefont {Barredo}}, \bibinfo
  {author} {\bibfnamefont {S.}~\bibnamefont {Weber}}, \bibinfo {author}
  {\bibfnamefont {N.}~\bibnamefont {Lang}}, \bibinfo {author} {\bibfnamefont
  {H.~P.}\ \bibnamefont {B{\"u}chler}}, \bibinfo {author} {\bibfnamefont
  {T.}~\bibnamefont {Lahaye}},\ and\ \bibinfo {author} {\bibfnamefont
  {A.}~\bibnamefont {Browaeys}},\ }\bibfield  {title} {\bibinfo {title}
  {Observation of a symmetry-protected topological phase of interacting bosons
  with rydberg atoms},\ }\href {https://doi.org/10.1126/science.aav9105}
  {\bibfield  {journal} {\bibinfo  {journal} {Science}\ }\textbf {\bibinfo
  {volume} {365}},\ \bibinfo {pages} {775} (\bibinfo {year}
  {2019})}\BibitemShut {NoStop}%
\bibitem [{\citenamefont {Weber}\ \emph {et~al.}(2018)\citenamefont {Weber},
  \citenamefont {de~L{\'{e}}s{\'{e}}leuc}, \citenamefont {Lienhard},
  \citenamefont {Barredo}, \citenamefont {Lahaye}, \citenamefont {Browaeys},\
  and\ \citenamefont {Büchler}}]{weber_18}%
  \BibitemOpen
  \bibfield  {author} {\bibinfo {author} {\bibfnamefont {S.}~\bibnamefont
  {Weber}}, \bibinfo {author} {\bibfnamefont {S.}~\bibnamefont
  {de~L{\'{e}}s{\'{e}}leuc}}, \bibinfo {author} {\bibfnamefont
  {V.}~\bibnamefont {Lienhard}}, \bibinfo {author} {\bibfnamefont
  {D.}~\bibnamefont {Barredo}}, \bibinfo {author} {\bibfnamefont
  {T.}~\bibnamefont {Lahaye}}, \bibinfo {author} {\bibfnamefont
  {A.}~\bibnamefont {Browaeys}},\ and\ \bibinfo {author} {\bibfnamefont
  {H.~P.}\ \bibnamefont {Büchler}},\ }\bibfield  {title} {\bibinfo {title}
  {Topologically protected edge states in small rydberg systems},\ }\href
  {https://doi.org/10.1088/2058-9565/aaca47} {\bibfield  {journal} {\bibinfo
  {journal} {Quantum Science and Technology}\ }\textbf {\bibinfo {volume}
  {3}},\ \bibinfo {pages} {044001} (\bibinfo {year} {2018})}\BibitemShut
  {NoStop}%
\bibitem [{\citenamefont {Weber}\ \emph {et~al.}(2022)\citenamefont {Weber},
  \citenamefont {Bai}, \citenamefont {Makki}, \citenamefont {M\"ogerle},
  \citenamefont {Lahaye}, \citenamefont {Browaeys}, \citenamefont {Daghofer},
  \citenamefont {Lang},\ and\ \citenamefont {B\"uchler}}]{weber_22}%
  \BibitemOpen
  \bibfield  {author} {\bibinfo {author} {\bibfnamefont {S.}~\bibnamefont
  {Weber}}, \bibinfo {author} {\bibfnamefont {R.}~\bibnamefont {Bai}}, \bibinfo
  {author} {\bibfnamefont {N.}~\bibnamefont {Makki}}, \bibinfo {author}
  {\bibfnamefont {J.}~\bibnamefont {M\"ogerle}}, \bibinfo {author}
  {\bibfnamefont {T.}~\bibnamefont {Lahaye}}, \bibinfo {author} {\bibfnamefont
  {A.}~\bibnamefont {Browaeys}}, \bibinfo {author} {\bibfnamefont
  {M.}~\bibnamefont {Daghofer}}, \bibinfo {author} {\bibfnamefont
  {N.}~\bibnamefont {Lang}},\ and\ \bibinfo {author} {\bibfnamefont {H.~P.}\
  \bibnamefont {B\"uchler}},\ }\bibfield  {title} {\bibinfo {title}
  {Experimentally accessible scheme for a fractional chern insulator in rydberg
  atoms},\ }\href {https://doi.org/10.1103/PRXQuantum.3.030302} {\bibfield
  {journal} {\bibinfo  {journal} {PRX Quantum}\ }\textbf {\bibinfo {volume}
  {3}},\ \bibinfo {pages} {030302} (\bibinfo {year} {2022})}\BibitemShut
  {NoStop}%
\bibitem [{\citenamefont {Semeghini}\ \emph {et~al.}(2021)\citenamefont
  {Semeghini}, \citenamefont {Levine}, \citenamefont {Keesling}, \citenamefont
  {Ebadi}, \citenamefont {Wang}, \citenamefont {Bluvstein}, \citenamefont
  {Verresen}, \citenamefont {Pichler}, \citenamefont {Kalinowski},
  \citenamefont {Samajdar}, \citenamefont {Omran}, \citenamefont {Sachdev},
  \citenamefont {Vishwanath}, \citenamefont {Greiner}, \citenamefont
  {Vuletić},\ and\ \citenamefont {Lukin}}]{semeghini_21}%
  \BibitemOpen
  \bibfield  {author} {\bibinfo {author} {\bibfnamefont {G.}~\bibnamefont
  {Semeghini}}, \bibinfo {author} {\bibfnamefont {H.}~\bibnamefont {Levine}},
  \bibinfo {author} {\bibfnamefont {A.}~\bibnamefont {Keesling}}, \bibinfo
  {author} {\bibfnamefont {S.}~\bibnamefont {Ebadi}}, \bibinfo {author}
  {\bibfnamefont {T.~T.}\ \bibnamefont {Wang}}, \bibinfo {author}
  {\bibfnamefont {D.}~\bibnamefont {Bluvstein}}, \bibinfo {author}
  {\bibfnamefont {R.}~\bibnamefont {Verresen}}, \bibinfo {author}
  {\bibfnamefont {H.}~\bibnamefont {Pichler}}, \bibinfo {author} {\bibfnamefont
  {M.}~\bibnamefont {Kalinowski}}, \bibinfo {author} {\bibfnamefont
  {R.}~\bibnamefont {Samajdar}}, \bibinfo {author} {\bibfnamefont
  {A.}~\bibnamefont {Omran}}, \bibinfo {author} {\bibfnamefont
  {S.}~\bibnamefont {Sachdev}}, \bibinfo {author} {\bibfnamefont
  {A.}~\bibnamefont {Vishwanath}}, \bibinfo {author} {\bibfnamefont
  {M.}~\bibnamefont {Greiner}}, \bibinfo {author} {\bibfnamefont
  {V.}~\bibnamefont {Vuletić}},\ and\ \bibinfo {author} {\bibfnamefont
  {M.~D.}\ \bibnamefont {Lukin}},\ }\bibfield  {title} {\bibinfo {title}
  {Probing topological spin liquids on a programmable quantum simulator},\
  }\href {https://doi.org/10.1126/science.abi8794} {\bibfield  {journal}
  {\bibinfo  {journal} {Science}\ }\textbf {\bibinfo {volume} {374}},\ \bibinfo
  {pages} {1242} (\bibinfo {year} {2021})}\BibitemShut {NoStop}%
\bibitem [{\citenamefont {Bai}\ \emph {et~al.}(2018)\citenamefont {Bai},
  \citenamefont {Bandyopadhyay}, \citenamefont {Pal}, \citenamefont {Suthar},\
  and\ \citenamefont {Angom}}]{bai_18}%
  \BibitemOpen
  \bibfield  {author} {\bibinfo {author} {\bibfnamefont {R.}~\bibnamefont
  {Bai}}, \bibinfo {author} {\bibfnamefont {S.}~\bibnamefont {Bandyopadhyay}},
  \bibinfo {author} {\bibfnamefont {S.}~\bibnamefont {Pal}}, \bibinfo {author}
  {\bibfnamefont {K.}~\bibnamefont {Suthar}},\ and\ \bibinfo {author}
  {\bibfnamefont {D.}~\bibnamefont {Angom}},\ }\bibfield  {title} {\bibinfo
  {title} {Bosonic quantum hall states in single-layer two-dimensional optical
  lattices},\ }\href {https://doi.org/10.1103/PhysRevA.98.023606} {\bibfield
  {journal} {\bibinfo  {journal} {Phys. Rev. A}\ }\textbf {\bibinfo {volume}
  {98}},\ \bibinfo {pages} {023606} (\bibinfo {year} {2018})}\BibitemShut
  {NoStop}%
\bibitem [{\citenamefont {Hafezi}\ \emph {et~al.}(2007)\citenamefont {Hafezi},
  \citenamefont {S\o{}rensen}, \citenamefont {Demler},\ and\ \citenamefont
  {Lukin}}]{hafezi_07}%
  \BibitemOpen
  \bibfield  {author} {\bibinfo {author} {\bibfnamefont {M.}~\bibnamefont
  {Hafezi}}, \bibinfo {author} {\bibfnamefont {A.~S.}\ \bibnamefont
  {S\o{}rensen}}, \bibinfo {author} {\bibfnamefont {E.}~\bibnamefont
  {Demler}},\ and\ \bibinfo {author} {\bibfnamefont {M.~D.}\ \bibnamefont
  {Lukin}},\ }\bibfield  {title} {\bibinfo {title} {Fractional quantum hall
  effect in optical lattices},\ }\href
  {https://doi.org/10.1103/PhysRevA.76.023613} {\bibfield  {journal} {\bibinfo
  {journal} {Phys. Rev. A}\ }\textbf {\bibinfo {volume} {76}},\ \bibinfo
  {pages} {023613} (\bibinfo {year} {2007})}\BibitemShut {NoStop}%
\bibitem [{\citenamefont {Palmer}\ and\ \citenamefont
  {Jaksch}(2006)}]{palmer_06}%
  \BibitemOpen
  \bibfield  {author} {\bibinfo {author} {\bibfnamefont {R.~N.}\ \bibnamefont
  {Palmer}}\ and\ \bibinfo {author} {\bibfnamefont {D.}~\bibnamefont
  {Jaksch}},\ }\bibfield  {title} {\bibinfo {title} {High-field fractional
  quantum hall effect in optical lattices},\ }\href
  {https://doi.org/10.1103/PhysRevLett.96.180407} {\bibfield  {journal}
  {\bibinfo  {journal} {Phys. Rev. Lett.}\ }\textbf {\bibinfo {volume} {96}},\
  \bibinfo {pages} {180407} (\bibinfo {year} {2006})}\BibitemShut {NoStop}%
\bibitem [{\citenamefont {Wang}\ \emph {et~al.}(2011)\citenamefont {Wang},
  \citenamefont {Gu}, \citenamefont {Gong},\ and\ \citenamefont
  {Sheng}}]{wang_11}%
  \BibitemOpen
  \bibfield  {author} {\bibinfo {author} {\bibfnamefont {Y.-F.}\ \bibnamefont
  {Wang}}, \bibinfo {author} {\bibfnamefont {Z.-C.}\ \bibnamefont {Gu}},
  \bibinfo {author} {\bibfnamefont {C.-D.}\ \bibnamefont {Gong}},\ and\
  \bibinfo {author} {\bibfnamefont {D.~N.}\ \bibnamefont {Sheng}},\ }\bibfield
  {title} {\bibinfo {title} {Fractional quantum hall effect of hard-core bosons
  in topological flat bands},\ }\href
  {https://doi.org/10.1103/PhysRevLett.107.146803} {\bibfield  {journal}
  {\bibinfo  {journal} {Phys. Rev. Lett.}\ }\textbf {\bibinfo {volume} {107}},\
  \bibinfo {pages} {146803} (\bibinfo {year} {2011})}\BibitemShut {NoStop}%
\bibitem [{\citenamefont {Gerster}\ \emph {et~al.}(2017)\citenamefont
  {Gerster}, \citenamefont {Rizzi}, \citenamefont {Silvi}, \citenamefont
  {Dalmonte},\ and\ \citenamefont {Montangero}}]{gerster_17}%
  \BibitemOpen
  \bibfield  {author} {\bibinfo {author} {\bibfnamefont {M.}~\bibnamefont
  {Gerster}}, \bibinfo {author} {\bibfnamefont {M.}~\bibnamefont {Rizzi}},
  \bibinfo {author} {\bibfnamefont {P.}~\bibnamefont {Silvi}}, \bibinfo
  {author} {\bibfnamefont {M.}~\bibnamefont {Dalmonte}},\ and\ \bibinfo
  {author} {\bibfnamefont {S.}~\bibnamefont {Montangero}},\ }\bibfield  {title}
  {\bibinfo {title} {Fractional quantum hall effect in the interacting
  hofstadter model via tensor networks},\ }\href
  {https://doi.org/10.1103/PhysRevB.96.195123} {\bibfield  {journal} {\bibinfo
  {journal} {Phys. Rev. B}\ }\textbf {\bibinfo {volume} {96}},\ \bibinfo
  {pages} {195123} (\bibinfo {year} {2017})}\BibitemShut {NoStop}%
\bibitem [{foo()}]{footnote}%
  \BibitemOpen
  \bibinfo {note} {{Note that strictly speaking it is not necessary to align
  lattice ${\rm B}$ so that, in the $y$ direction, sites of lattice ${\rm A}$
  are centered between sites of lattice ${\rm B}$, because the resulting
  complex hoppings along $x$ could be gauged away.}}\BibitemShut {Stop}%
\end{thebibliography}%

\end{document}